\documentclass[sigconf]{acmart}
\AtBeginDocument{%
  \providecommand\BibTeX{{%
    \normalfont B\kern-0.5em{\scshape i\kern-0.25em b}\kern-0.8em\TeX}}}
\copyrightyear{2022} 
\acmYear{2022} 
\setcopyright{acmcopyright}
\acmConference[MSR '22]{19th International Conference on Mining Software Repositories}{May 23--24, 2022}{Pittsburgh, PA, USA}
\acmBooktitle{19th International Conference on Mining Software Repositories (MSR '22), May 23--24, 2022, Pittsburgh, PA, USA}
\acmPrice{15.00}
\acmDOI{10.1145/3524842.3528520}
\acmISBN{978-1-4503-9303-4/22/05}

\usepackage[]{todonotes}

\begin{document}

\title{Bot Detection in GitHub Repositories}

\author{Natarajan Chidambaram}
\orcid{0000-0002-2295-8928}
\email{natarajan.chidambaram@umons.ac.be}
\affiliation{%
  \institution{Software Engineering Lab, University of Mons}
  \city{Mons}
  \country{Belgium}
}

\author{Pooya Rostami Mazrae}
\orcid{1234-1234-1234-1234}
\email{pooya.rostamimazrae@umons.ac.be}
\affiliation{%
  \institution{Software Engineering Lab, University of Mons}
  \city{Mons}
  \country{Belgium}
}

\begin{abstract}
Contemporary social coding platforms like GitHub promote collaborative development. Many open-source software repositories hosted in these platforms use machine accounts (bots) to automate and facilitate a wide range of effort-intensive and repetitive activities. Determining if an account corresponds to a bot or a human contributor is important for socio-technical development analytics, for example, to understand how humans collaborate and interact in the presence of bots, to assess the positive and negative impact of using bots, to identify the top project contributors, to identify potential bus factors, and so on. Our project aims to include the trained machine learning (ML) classifier from the BoDeGHa bot detection tool as a plugin to the GrimoireLab software development analytics platform.

In this work, we present the procedure to form a pipeline for retrieving contribution and contributor data using Perceval, distinguishing bots from humans using BoDeGHa, and visualising the results using Kibana.

\end{abstract}
\maketitle

\section{Introduction}
\label{sec:intro}
Social coding platforms like GitHub promote collaboration and interaction between developers \cite{software-social}. Along with this opportunity for engagement, developers also face some workload in performing error-prone, time-intensive or repetitive tasks such as conducting quality checks, testing, code reviewing, merging, building, deploying, and so on \cite{Wessel2018}. Thus, the developer overload increases as the frequency of these tasks increases \cite{contributions_count}. Bots (machine accounts that act with as minimal human intervention as possible) are therefore  frequently used to face this ever-increasing complexity in software development \cite{Erlenhov2019}.

There are many bot accounts that are built by software developers and are used only in a specific set of repositories. 
For instance, the \emph{highfive} account is responsible for greeting and assigning issues to contributors in \emph{servo/servo}, one of the largest packages distributed through the Cargo package manager. Similarly, the \emph{bors-diem} bot is managing the merging of pull requests (PR) in the \emph{diem/diem} Cargo package. 
However, they are not evidently identified as bots as they neither have \textit{bot} in their name nor their GitHub description/bio\footnote{\url{https://github.com/bors-servo} and \url{https://github.com/libra-action}} conveys this explicitly. The challenge of identifying bot accounts makes it difficult to conduct socio-technical analysis that need to distinguish human from bot behaviour, while doing so would be valuable for researchers to better understand the positive and negative impact of bots in software development, as well as for practitioners and organizations that want to accredit human project contributors~\cite{botse_paper_2}.
\smallskip

The GrimoireLab software development analytics toolkit\footnote{\url{https://chaoss.github.io/grimoirelab/}} provides \emph{data retrieval} capability with tools such as Perceval and Graal, 
\emph{data enrichment} using tools such as HatStall and SortingHat (for merging duplicate contributor identities), and \emph{data visualisation} using tools such as Kidash and KiBiter (based on Kibana dashboards for visualising Elasticsearch data).
Adding a bot identification model as one of GrimoireLab's data enrichment components would enable GrimoireLab users to collect data from GitHub repositories, identify the bot accounts, and visualise/analyse socio-technical collaborative development activities using a single platform.

\begin{figure}[!h]
	\centering
	\includegraphics[width = \columnwidth]{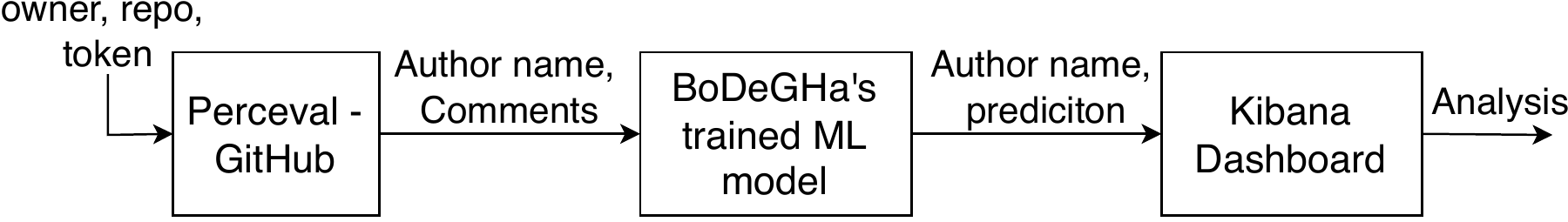}
	\caption{Integrating BoDeGHa in the GrimoireLab pipeline}
	\label{fig:pipeline}
\end{figure}

As part of the MSR 2021 Hackathon\footnote{Team: NAP; repository: \url{https://github.com/pooya-rostami/Hackathon-21}},
we propose the creation of such an end-to-end pipeline (illustrated in Figure~\ref{fig:pipeline}) that integrates BoDeGHa~\cite{Bodegha}, a tool to identify bots in GitHub repositories on the basis of their PR and issue commenting activities. BoDeGHa comes with a trained machine learning classifier that could be added as a part of the GrimoireLab pipeline.

\section{Approach and Preliminary Results}
\label{sec:PrelimResult}
Since we aim to improve GrimoireLab's bot detection capability by integrating BoDeGHa's trained machine learning classifier (hereafter addressed as BoDeGHa), we need to execute the pipeline of Figure~\ref{fig:pipeline} with a set of GitHub repositories. This section presents a step-by-step process that needs to be followed to query, predict and visualise the number of contributors along with their type.

In the first step, we pass the GitHub repository and owner name along with the GitHub API token as an argument to Perceval for querying issues and pull requests (PRs) in the corresponding repository. The tool returns the comments present in each issue and PR along with user- and repository-specific information. Next, we extract the fields \emph{issue/PR number}, \emph{comments}, \emph{created at}, and \emph{corresponding author names} and save them in a CSV file to execute BoDeGHa bot identification tool. The ML model predicts the type of contributor (\emph{bot} or \emph{human}) in the corresponding repository based on repetition and patterns in the comments made by the contributors in issues and PRs. The contributor names along with their predicted type (\emph{bot} or \emph{human}) are stored in a CSV file. In the last step, we use Kibana (an open user interface that is used for visualising Elasticsearch data) to present BoDeGHa's prediction results. 

\begin{figure}[!t]
	\centering
	\includegraphics[width = \columnwidth]{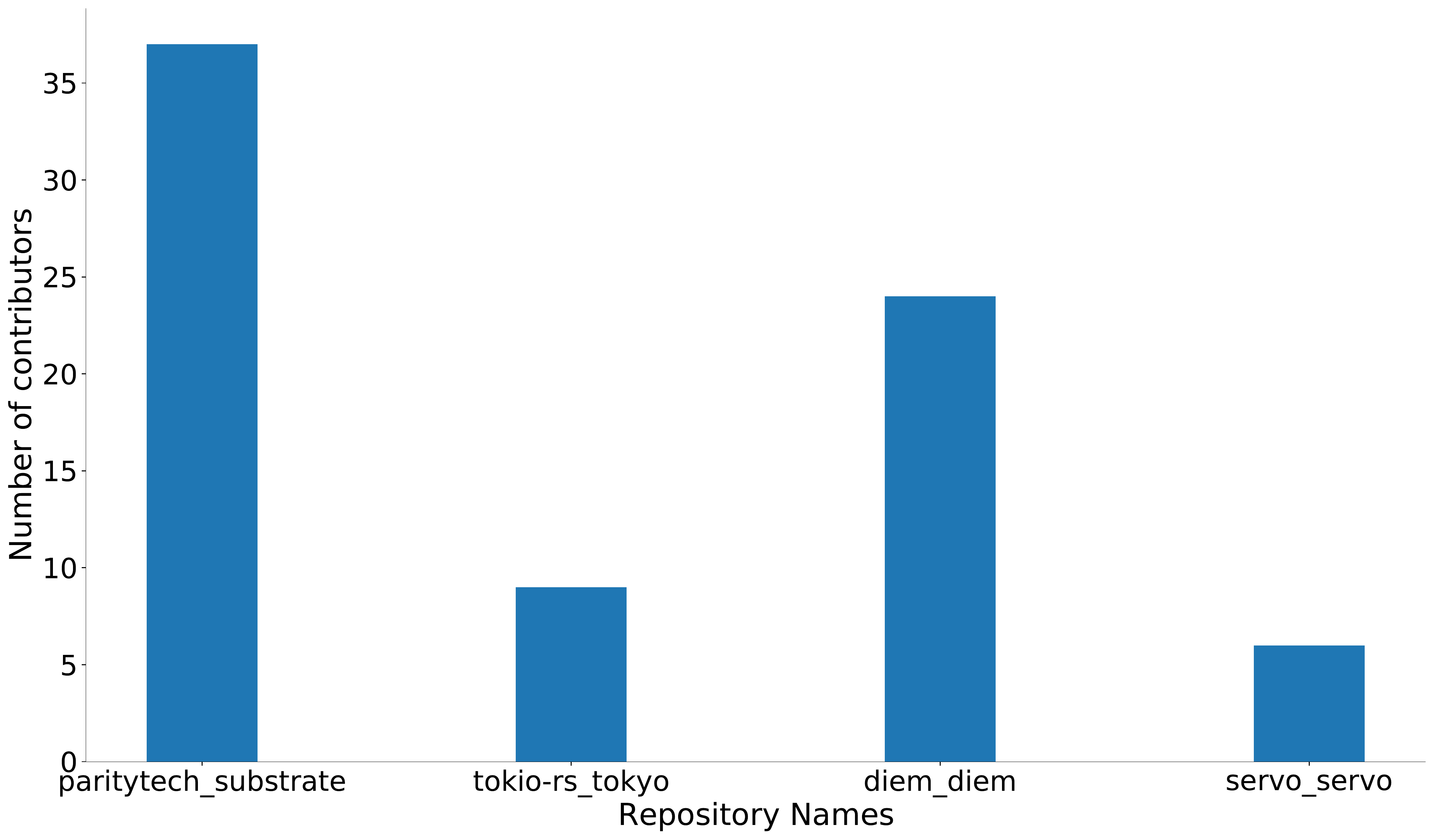}
	\caption{Number of contributors present in each repository}
	\label{fig:accounts}
\end{figure}

To illustrate this process of integrating BoDeGHa into the GrimoireLab pipeline, we use four large GitHub repositories of Cargo packages, namely, \emph{diem/diem}, \emph{servo/servo}, \emph{SergioBenitez/Rocket}, and \emph{paritytech/substrate}. Due to their size and popularity, these repositories are likely to use bots in their collaborative software development process.

\begin{figure}[!t]
	\centering
	\includegraphics[width = \columnwidth]{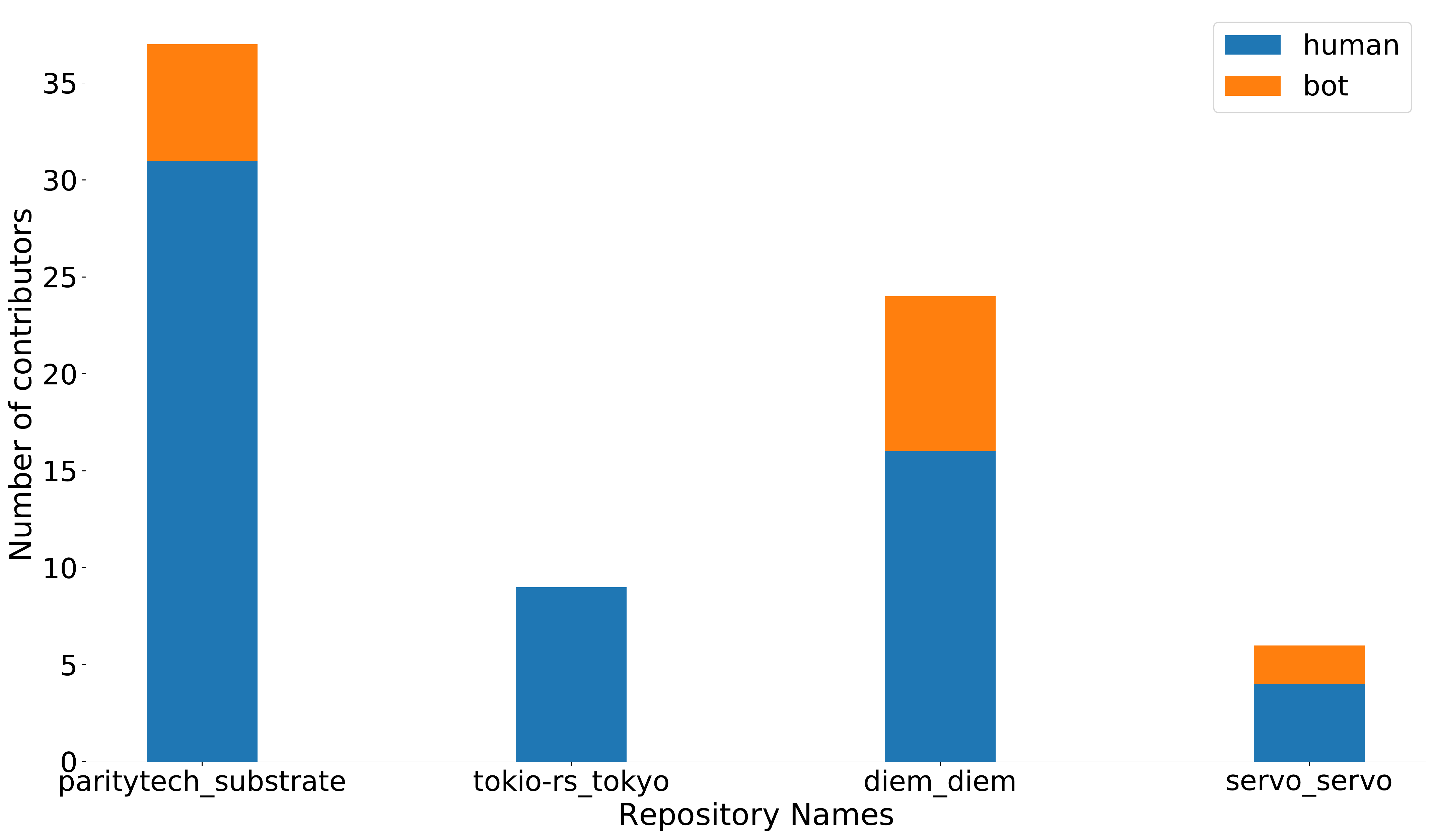}
	\caption{Number and type of contributors present in each repository}
	\label{fig:classification}
\end{figure}

Figure~\ref{fig:accounts} shows, without using the integrated bot prediction tool, the number of contributors that were actively posting comments in each repository between December 2021 and January 2022. 
In contrast, Figure~\ref{fig:classification} provides a fine-grained view by taking into account the type of contributor (\emph{human} or \emph{bot}) in each repository. We observe a proportionally high number of bots in three of the considered repositories:  6 out of 37 contributors are bots in \emph{paritytech/substrate}, 8 out of 24 in \emph{diem/diem} and 2 out of 6 in \emph{servo/servo}. Also, it can be observed that all the issue and PR comments in \emph{tokio-rs/tokio} were made by human contributors. 

\section{Going Further}

Knowing the types of contributors within a repository could be leveraged for many other types of visualisation through the Kibana dashboard. For example, it would be useful to come up with visualisations that reveal whether bots are among the most active contributors within a given software repository. Similarly, one could provide visualisations involving data from multiple repositories, in order to understand how widespread specific bots are being used.

Given that BoDeGHa can occasionally produce incorrect classifications, it would also be useful to provide an interface in the BoDeGHa plugin to allow an operator to rectify misclassified contributor types (in a CSV file that has the predictions) before visualising the results.

In a similar way, the BoDeGHa plugin could be integrated with SortingHat, GrimoireLab's component in charge of managing contributors' data (including the flag to mark them as bot) and of adding contributors' details in the data shown in Kibiter \cite{Moreno2019}.

\section{Conclusion}
\label{sec:Conclusion}

Bots are being prevalently used in collaborative software development on GitHub to face ever-increasing complexity, by automating effort-intensive and repetitive activities. Identifying these bots is valuable for researchers performing socio-technical analysis in software development, as well as for practitioners and organizations that want to accredit human contributors. BoDeGHa is one such tool that identifies bot accounts based on their commenting activity in issues and pull requests.
As part of the MSR 2022 Hackathon, we integrated BoDeGHa's trained machine learning classifier into the GrimoireLab pipeline, by extracting repository data with Perceval, identifying bot contributors with BoDeGHa, and visualising the results with Kibana dashboard.

\begin{acks}
This work is supported by DigitalWallonia4.AI research project ARIAC (grant number 2010235), as well as by the ARC-21/25 UMONS3 Action de Recherche Concertée financée par le Ministère de la Communauté française – Direction générale de l’Enseignement non obligatoire et de la Recherche scientifique
\end{acks}

\newpage
\bibliographystyle{ACM-Reference-Format}
\bibliography{MSR2022}


\begin{thebibliography}{7}

%%% the Web address.

\ifx \showCODEN    \undefined \def \showCODEN     #1{\unskip}     \fi
\ifx \showDOI      \undefined \def \showDOI       #1{#1}\fi
\ifx \showISBNx    \undefined \def \showISBNx     #1{\unskip}     \fi
\ifx \showISBNxiii \undefined \def \showISBNxiii  #1{\unskip}     \fi
\ifx \showISSN     \undefined \def \showISSN      #1{\unskip}     \fi
\ifx \showLCCN     \undefined \def \showLCCN      #1{\unskip}     \fi
\ifx \shownote     \undefined \def \shownote      #1{#1}          \fi
\ifx \showarticletitle \undefined \def \showarticletitle #1{#1}   \fi
\ifx \showURL      \undefined \def \showURL       {\relax}        \fi
% invisible to TeX
\providecommand\bibfield[2]{#2}
\providecommand\bibinfo[2]{#2}
\providecommand\natexlab[1]{#1}
\providecommand\showeprint[2][]{arXiv:#2}

\bibitem[Dabbish et~al\mbox{.}(2012)]%
        {software-social}
\bibfield{author}{\bibinfo{person}{Laura Dabbish}, \bibinfo{person}{Colleen
  Stuart}, \bibinfo{person}{Jason Tsay}, {and} \bibinfo{person}{Jim Herbsleb}.}
  \bibinfo{year}{2012}\natexlab{}.
\newblock \showarticletitle{Social Coding in {GitHub}: Transparency and
  Collaboration in an Open Software Repository}. In
  \bibinfo{booktitle}{\emph{International Conference on Computer Supported
  Cooperative Work}}. \bibinfo{pages}{1277--1286}.
\newblock
\urldef\tempurl%
\url{https://doi.org/10.1145/2145204.2145396}
\showDOI{\tempurl}


\bibitem[Erlenhov et~al\mbox{.}(2019)]%
        {Erlenhov2019}
\bibfield{author}{\bibinfo{person}{Linda Erlenhov}, \bibinfo{person}{Francisco
  Gomes~de Oliveira~Neto}, \bibinfo{person}{Riccardo Scandariato}, {and}
  \bibinfo{person}{Philipp Leitner}.} \bibinfo{year}{2019}\natexlab{}.
\newblock \showarticletitle{Current and Future Bots in Software Development}.
  In \bibinfo{booktitle}{\emph{International Workshop on Bots in Software
  Engineering (BotSE)}}. IEEE, \bibinfo{pages}{7--11}.
\newblock
\urldef\tempurl%
\url{https://doi.org/10.1109/BotSE.2019.00009}
\showDOI{\tempurl}


\bibitem[Golzadeh et~al\mbox{.}(2022)]%
        {botse_paper_2}
\bibfield{author}{\bibinfo{person}{Mehdi Golzadeh}, \bibinfo{person}{Alexandre
  Decan}, {and} \bibinfo{person}{Natarajan Chidambaram}.}
  \bibinfo{year}{2022}\natexlab{}.
\newblock \showarticletitle{On the accuracy of bot detection techniques}. In
  \bibinfo{booktitle}{\emph{International Workshop on Bots in Software
  Engineering (BotSE)}}. IEEE.
\newblock


\bibitem[Golzadeh et~al\mbox{.}(2021)]%
        {Bodegha}
\bibfield{author}{\bibinfo{person}{Mehdi Golzadeh}, \bibinfo{person}{Alexandre
  Decan}, \bibinfo{person}{Damien Legay}, {and} \bibinfo{person}{T. Mens}.}
  \bibinfo{year}{2021}\natexlab{}.
\newblock \showarticletitle{A ground-truth dataset and classification model for
  detecting bots in GitHub issue and PR comments}.
\newblock \bibinfo{journal}{\emph{Journal of Systems and Software}}
  \bibinfo{volume}{175} (\bibinfo{year}{2021}).
\newblock
\urldef\tempurl%
\url{https://doi.org/10.1016/j.jss.2021.110911}
\showDOI{\tempurl}


\bibitem[Moreno et~al\mbox{.}(2019)]%
        {Moreno2019}
\bibfield{author}{\bibinfo{person}{David Moreno}, \bibinfo{person}{Santiago
  Due{\~n}as}, \bibinfo{person}{Valerio Cosentino},
  \bibinfo{person}{Miguel~Angel Fernandez}, \bibinfo{person}{Ahmed Zerouali},
  \bibinfo{person}{Gregorio Robles}, {and} \bibinfo{person}{Jesus~M.
  Gonzalez-Barahona}.} \bibinfo{year}{2019}\natexlab{}.
\newblock \showarticletitle{SortingHat: Wizardry on Software Project Members}.
  In \bibinfo{booktitle}{\emph{International Conference on Software
  Engineering: Companion Proceedings (ICSE-Companion)}}. IEEE,
  \bibinfo{pages}{51--54}.
\newblock
\urldef\tempurl%
\url{https://doi.org/10.1109/ICSE-Companion.2019.00036}
\showDOI{\tempurl}


\bibitem[Wessel et~al\mbox{.}(2018)]%
        {Wessel2018}
\bibfield{author}{\bibinfo{person}{Mairieli Wessel},
  \bibinfo{person}{Bruno~Mendes {De Souza}}, \bibinfo{person}{Igor
  Steinmacher}, \bibinfo{person}{Igor~S. Wiese}, \bibinfo{person}{Ivanilton
  Polato}, \bibinfo{person}{Ana~Paula Chaves}, {and} \bibinfo{person}{Marco~A.
  Gerosa}.} \bibinfo{year}{2018}\natexlab{}.
\newblock \showarticletitle{The power of bots: Understanding bots in {OSS}
  projects}.
\newblock \bibinfo{journal}{\emph{The ACM International Conference on
  Human-Computer Interaction}} (\bibinfo{year}{2018}).
\newblock
\showISSN{25730142}
\urldef\tempurl%
\url{https://doi.org/10.1145/3274451}
\showDOI{\tempurl}


\bibitem[Young et~al\mbox{.}(2021)]%
        {contributions_count}
\bibfield{author}{\bibinfo{person}{Jean-Gabriel Young}, \bibinfo{person}{Amanda
  Casari}, \bibinfo{person}{Katie McLaughlin}, \bibinfo{person}{Milo~Z.
  Trujillo}, \bibinfo{person}{Laurent H{\'e}bert-Dufresne}, {and}
  \bibinfo{person}{James~P. Bagrow}.} \bibinfo{year}{2021}\natexlab{}.
\newblock \showarticletitle{Which contributions count? Analysis of attribution
  in open source}.
\newblock \bibinfo{journal}{\emph{International Conference on Mining Software
  Repositories (MSR)}} (\bibinfo{year}{2021}), \bibinfo{pages}{242--253}.
\newblock
\urldef\tempurl%
\url{https://doi.org/10.1109/MSR52588.2021.00036}
\showDOI{\tempurl}


\end{thebibliography}

\end{document}